\documentclass[runningheads]{llncs}
\usepackage{dialogue}
\usepackage{tabularx}
\usepackage{amssymb}
\usepackage{amsmath} 
\usepackage{algorithm}
\usepackage{algpseudocode}
\usepackage{float}
\usepackage{mdframed}
\usepackage{xcolor}
\usepackage{xurl}
\usepackage{hyperref}
\usepackage{url}
\usepackage{threeparttable}
\usepackage[T1]{fontenc}
\usepackage{graphicx}
\usepackage{enumitem}
\usepackage{booktabs}
\usepackage{makecell}
\usepackage{subcaption}
\usepackage{array}
\usepackage{orcidlink}
\begin{document}
\title{Evaluating a Data-Driven Redesign Process for Intelligent Tutoring Systems}
\titlerunning{Data-Driven Redesign Process Evaluation for Intelligent Tutoring Systems}
\author{Qianru Lyu\inst{1}\textsuperscript{*}\orcidlink{0000-0003-0650-1570} \and 
Conrad Borchers\inst{1}\textsuperscript{*}\orcidlink{0000-0003-3437-8979} \and
Meng Xia\inst{2}\orcidlink{0000-0002-2676-9032} \and
Karen Xiao\inst{3}\orcidlink{0009-0001-9717-6111} \and
Paulo F. Carvalho\inst{1}\orcidlink{0000-0002-0449-3733} \and
Kenneth R. Koedinger\inst{1}\orcidlink{0000-0002-5850-4768} \and
Vincent Aleven\inst{1}\orcidlink{0000-0002-1581-6657}}
\authorrunning{Lyu et al.}
\institute{Carnegie Mellon University\\
\email{\{qlyu,cborcher,pcarvalh,krk,aleven\}@cs.cmu.edu}\\
\and
Texas A\&M University\\
\email{mengxia@tamu.edu}
\and
Wellesley College\\
\email{kx100@wellesley.edu}\\
*Equal contribution.
}
\maketitle 
\begin{abstract}
Past research has defined a general process for the data-driven redesign of educational technologies and has shown that in carefully-selected instances, this process can help make systems more effective. In the current work, we test the generality of the approach by applying it to four units of a middle-school mathematics intelligent tutoring system that were selected not based on suitability for redesign, as in previous work, but on topic. We tested whether the redesigned system was more effective than the original in a classroom study with 123 students. Although the learning gains did not differ between the conditions, students who used the Redesigned Tutor had more productive time-on-task, a larger number of skills practiced, and greater total knowledge mastery. The findings highlight the promise of data-driven redesign even when applied to instructional units \textit{not} selected as likely to yield improvement, as evidence of the generality and wide applicability of the method. 
\end{abstract}


\section{Introduction and Related Work} 

Iterative, data-driven redesign promises to improve AIED systems but is not widespread \cite{Koedinger2013NewPotentials,huang2021general}. Even carefully designed instructional systems are rarely optimal at first deployment \cite{Koedinger2013NewPotentials}. Data-driven redesign uses data from prior use to inform targeted revisions, which has been shown to lead to improvement of educational technologies~\cite{Bakharia2016LearningDesignAnalytics,Liu2017ExplanatoryModels}. Intelligent tutoring systems (ITS) are particularly well-suited to this approach because they generate rich log data. %

Although many data mining methods improve the prediction of student performance, few drive interventions with measurable effects on learners \cite{Clow2012LearningAnalyticsCycle}. Prior research has created a redesign strategy, focused on the refinement of knowledge component (KC) models, that has repeatedly yielded benefits in past studies \cite{Koedinger2013NewPotentials,Koedinger2013StudentModels,huang2021general}. A KC model decomposes a domain into fine-grained skills. KC models can guide instructional design in many ways, including the design of problem sets (i.e., practice items), the design of scaffolding within problems, tracking knowledge growth for mastery learning \cite{long2018exactly}. Many core functions of an ITS depend on an accurate KC model. Structuring instruction around a well-specified KC model can substantially improve learning by aligning practice with the cognitive operations required for problem solving \cite{Lovett2008OLIStatistics}. However, KC models often contain deficiencies that undermine their effectiveness when used in instruction.
Prior work has shown that student performance data can be used to diagnose deficiencies in KC models and guide systematic redesign \cite{aleven2025instruction,Koedinger2013StudentModels}. These works provided ``close-the-loop'' evidence linking data-driven KC refinement to pedagogical effectiveness, a connection we extend in the present study. Additional close-the-loop studies reporting learning improvements from data-driven redesign include \cite{koedinger2016closing,Liu2017ExplanatoryModels}, as well as related syntheses and applied redesign efforts \cite{Koedinger2013NewPotentials,Lovett2008OLIStatistics}.

Building on this work, Huang et al. \cite{huang2021general} introduced the General Multi-method Approach to Data-Driven Redesign, which integrates KC model refinement, instructional redesign, and optimization of individualized learning. The approach addresses misallocated practice across KCs, insufficient scaffolding for hard KCs identified by persistently high learning curves, and recurring errors. It combines newly proposed methods, such as Difficulty Factors Effect Analysis (to identify difficulty factors), with established techniques from educational data mining, many supported by DataShop \cite{Stamper2011DataShop}. A central strategy in Huang et al. \cite{huang2021general} was focused practice, which directs additional practice to difficult KCs while avoiding redundant practice on easier or untargeted skills. In a one-month high school classroom study, the Redesigned Tutor produced significantly higher learning outcomes than the original while reducing both over- and underpractice \cite{huang2021general}.

Despite these promising results, evidence for the generality of these approaches is limited. The redesign process proposed by Huang et al. \cite{huang2021general} has been evaluated on a narrow range of middle-school mathematics content, leaving open whether it reliably improves learning when applied to new instructional units. Assessing generality requires applying the process across additional content areas and instructional contexts. Moreover, prior evaluations have focused largely on aggregate learning gains, providing limited insight into how redesign affects students’ learning processes during practice. To address these gaps, the present study applies the data-driven redesign process to multiple new instructional units and evaluates both learning outcomes and practice efficiency at the process level. Three questions are explored in this study: 

\textbf{RQ1:} How does Huang et al.~\cite{huang2021general}'s redesign process, applied to an ITS for middle-school mathematics, reveal tutor limitations and help generate redesign ideas?
\textbf{RQ2:} How does the Redesigned Tutor affect students' learning \textit{outcome}, compared to the Original Tutor? \textbf{RQ3:} How does the Redesigned Tutor affect students' learning \textit{process}, compared to the Original Tutor?

We examined a redesign of MathTutor, an ITS shown to effectively support middle-school mathematics learning \cite{borchers2023makes}. MathTutor implements the core ITS features described above. We redesigned four of its 65 problem sets: graph interpretation, plotting, and two equation-solving units. These units have also undergone multiple rounds of refinement based on classroom use and teacher feedback, though not through data-driven redesign. Hence, they constitute a strong test case for evaluating the potential impact of a systematic redesign process.

\section{Data-Driven Redesign Process (RQ1)} 
\label{sec:redesign-process-general}

We used existing log data from four MathTutor units (including 103 students, 22,529 transactions; 50.19 practice hours) for the data-driven redesign of these units. We followed a three-step data-driven redesign process \cite{huang2021general}. Table \ref{tab:unit_applications} summarizes how the three redesign goals were instantiated across units.
First, we examined learning curves in DataShop \cite{Stamper2011DataShop} to identify ill-specified KCs or KCs ineffectively supported and redefined the KC model. Candidate KC models were evaluated by whether they exhibited gradual learning, reflected in smooth learning curves and good AFM fit \cite{Liu2017ExplanatoryModels}. Applying this step yielded different results in each unit; see Table \ref{tab:unit_applications}. In Unit 7.05 (Graph Interpretation), some problems required students to infer coordinates between labeled grid points, yet the tutor treated these as simple lookup tasks; learning-curve analysis, therefore, revealed an unmodeled interpolation skill, motivating KC splitting for inferring coordinates with and without interpolation and adding two skills. In Unit 7.06, model comparison did not support refinement. For equation solving, operations involving negative terms (e.g., subtracting -3 or interpreting ``-x'' as ``-1x'') are more difficult for students, but were previously modeled as equivalent to positive cases; separating these sign-handling skills \cite{long2018exactly} improved fit and parsimony (AIC).

\begin{figure}[htpb]
 \centering
 \begin{subfigure}[t]{0.48\linewidth}
  \centering
  \includegraphics[width=\linewidth]{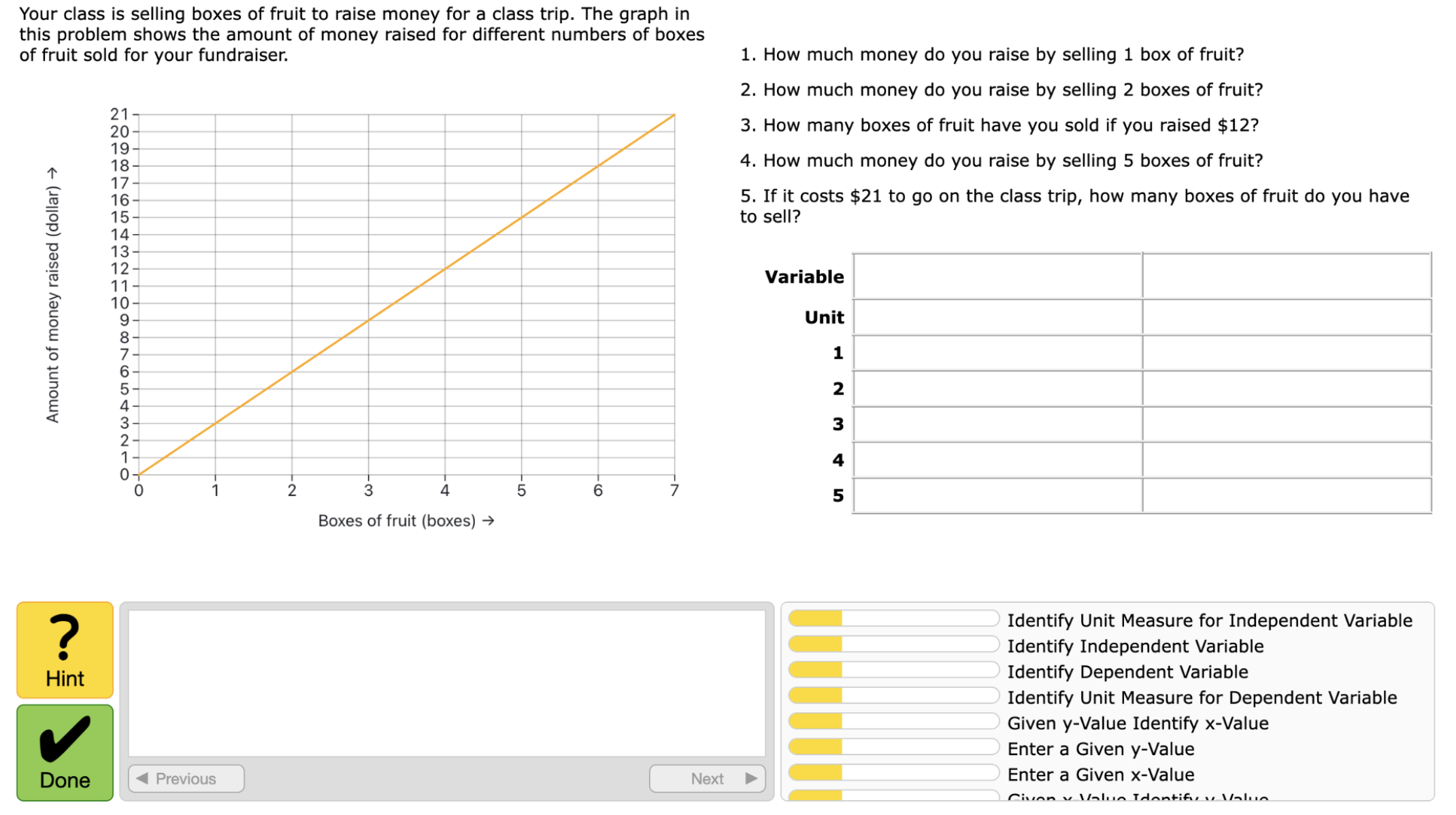}
  \caption{}
  \label{fig:exp2a}
 \end{subfigure}
 \hfill
 \begin{subfigure}[t]{0.48\linewidth}
  \centering
  \includegraphics[width=\linewidth]{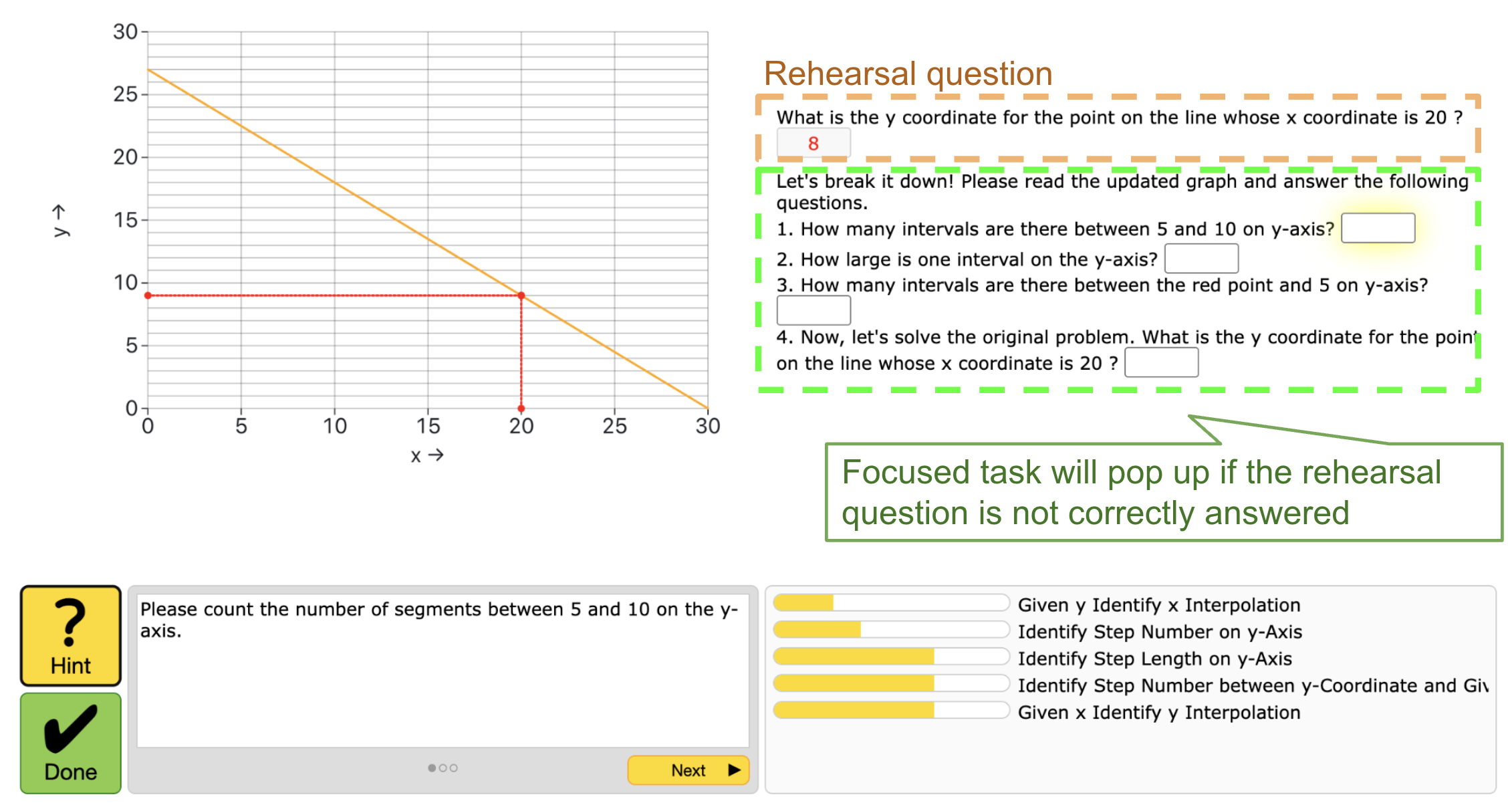}
  \caption{}
    \label{fig:exp2b}
  \end{subfigure}
  \caption{Original (left) and redesigned (right) Unit 7.05 Graph Interpretation tutor. The redesigned version begins with an interpolation rehearsal and conditionally reveals focused tasks when the rehearsal is answered incorrectly.}
  \label{fig:exp2}
\end{figure}

Second, we redesigned the tutor units to better scaffold and practice challenging or newly identified KCs with changes to the problem types, scaffolding, and problem selection algorithm (Table \ref{tab:unit_applications}). We designed new tasks for targeted practice of difficult KCs (``focused part tasks'' \cite{huang2021general}) that isolate a KC for efficient practice and dynamically replace whole tasks when mastery is lacking, reducing overpractice. For example, in Unit 7.05, we created new tasks that scaffold the reasoning steps needed for interpolating when reading a point off of a graph (see Fig. \ref{fig:exp2}). All redesigns were implemented in CTAT \cite{aleven2025integrated}. 

Third, we implemented an adaptive problem selection algorithm \cite{huang2021general}, which prioritizes unmastered KCs, reduces redundant practice on mastered skills, and allocates practice more efficiently. Offline simulation was conducted to check under-practice and over-practice of specific KCs. As a result, we added problems to the equation-solving unit to ensure sufficient practice for newly split KCs, including negative subtraction, negative division, and division with ``-x.'' We also enabled \textit{Mastered Step Skipping} \cite{xia2025optimizing}, which automatically completes remaining steps once all relevant KCs are estimated as mastered at a 95\% threshold. With this mechanism, the student would only need to reduce more advanced equations (e.g., those involving variables on both sides or composite terms in parentheses) to a form that was already familiar (e.g., two-step equations), without having to solve that familiar form, thus avoiding over-practice.

\begin{table}[htpb]
\centering
\renewcommand{\arraystretch}{1.2}
\caption{Unit-specific realization of redesign goals.}
\label{tab:unit_applications}
\resizebox{\linewidth}{!}{%
\begin{tabular}{
>{\raggedright\arraybackslash}p{2.3cm}|
>{\raggedright\arraybackslash}p{4cm}|
>{\raggedright\arraybackslash}p{4cm}|
>{\raggedright\arraybackslash}p{3.6cm}
}
\toprule
\textbf{Goal} & \textbf{Unit 7.05} & \textbf{Unit 7.06} & \textbf{Equation Solving I, II} \\
\midrule
\textbf{Goal 1: Discoveries made in data}&
Identified a hidden interpolation skill; split graph-reading KCs based on AFM evidence. &
The \emph{draw-a-line} KC remained highly difficult; no KC refinement supported. &
Separated sign-handling cases (positive, negative, and $-x$), building on Long et~al.~\cite{long2018exactly}. \\
\hline
\textbf{Goal 2: Tutor redesign} &
Added focused interpolation tasks; revised wording; improved graph clarity and scaffolding. &
Relaxed step-order constraints; refined hints; added focused graphing tasks with pre-filled tables. &
Updated hints and scaffolds to align with refined KCs; added adaptive step support. \\
\hline
\textbf{Goal 3: Adaptive support} &
Adaptive problem selection following Huang et~al.~\cite{huang2023supporting}. &
Adaptive problem selection following Huang et~al.~\cite{huang2023supporting}. &
Adaptive problem selection~\cite{huang2023supporting}, mastered step skipping~\cite{xia2025optimizing}. \\

\bottomrule
\end{tabular}
}
\end{table}

\section{Classroom Evaluation (RQ2 \& RQ3)} 

\subsection{Participants and Procedure}
A total of 123 students from 5 middle schools in the U.S. participated in the study. The sample spanned 8 classes from grades 7 and 8, taught by six different teachers. The study was conducted during students’ regularly scheduled mathematics periods and followed an approved IRB protocol. All data from the study are available in DataShop (datasets \#5868 and \#6283) \cite{Stamper2011DataShop}. The study design and primary analyses were preregistered prior to data analysis (\url{https://osf.io/xh584/}).

\begin{figure}
  \centering
  \includegraphics[width=0.75\linewidth]{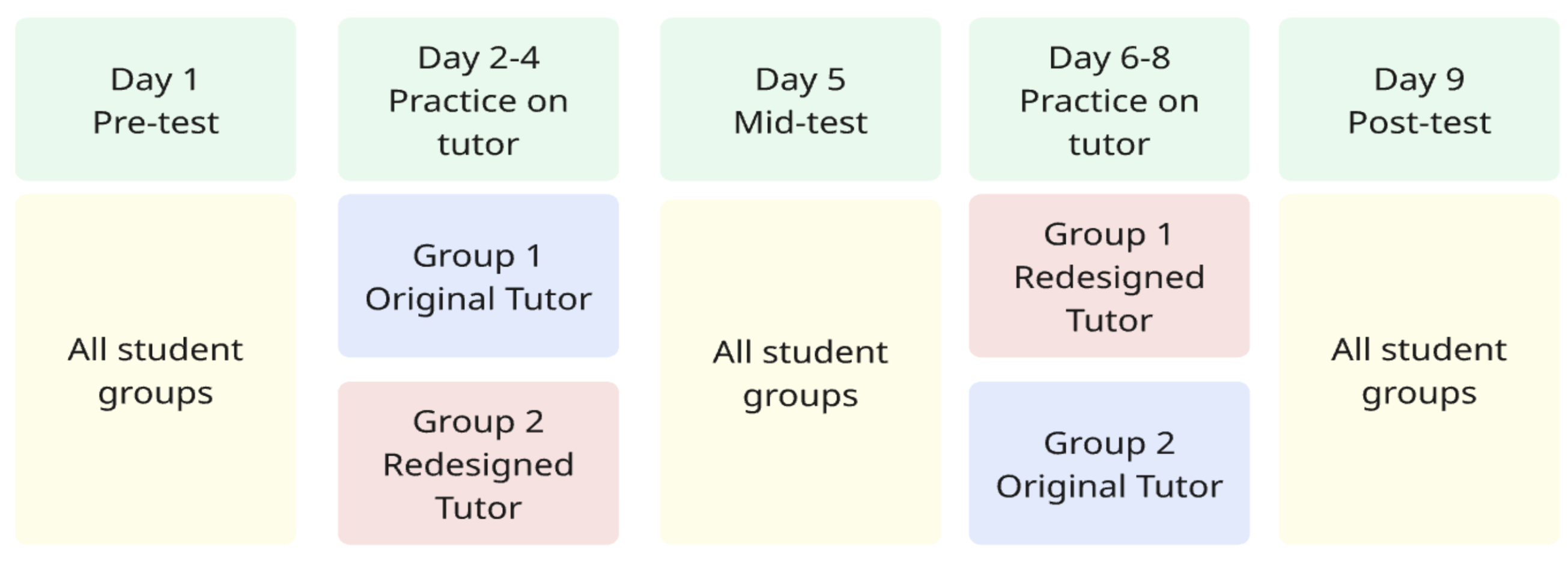}
  \caption{Experimental design. The crossover happened between Days 4 and 6.}
  \label{fig:exp-design}
\end{figure}

The study followed a randomized within-subjects crossover design with two tutor conditions and pre-, mid-, and post-tests to measure learning gains (Fig. \ref{fig:exp-design}). Students completed a pre-test on Day 1, practiced with a tutor on Days 2–4, took a mid-test on Day 5, practiced again on Days 6–8, and completed a post-test on Day 9. During the first practice phase, students were randomly assigned to either the Original or Redesigned Tutor; after the mid-test, the groups switched conditions. Each practice session lasted approximately 20–25 minutes. Two isomorphic test forms (A and B) and test order were counterbalanced across conditions, with about 20 minutes allotted per test. All tutoring features, including hints, feedback, and skill bars, were disabled during testing.

\subsection{Data Analysis}

\subsubsection{Learning Gains (RQ2)}
To examine whether the Redesigned Tutor affected learning gains (post minus pre) relative to the Original Tutor (RQ2), we analyzed pre--post test performance using a linear mixed-effects model. The model included fixed effects of \textit{condition} (Original vs. Redesigned), \textit{test time} (Pre vs. Post), and their interaction (the differential gain by condition). Random intercepts for student, school, and content unit accounted for the hierarchical structure of the data. Students missing either pre- or post-test scores were excluded (final $N=95$). Model diagnostics indicated model assumptions were tenable. All mixed-model tests were conducted with the \texttt{lmerTest} package in R \cite{kuznetsova2017lmertest}.

\subsubsection{Learning Processes and Within-Tutor Learning (RQ3)}
\label{sec:methods:process-analysis}

To address RQ3, we analyzed log data to compare student practice behavior across the two tutor versions. Time-on-task was computed by subtracting idle periods, defined as more than two minutes of inactivity following \cite{holstein2018student}, from total logged-in time, yielding a measure of active engagement. To estimate within-tutor learning, we derived end-of-practice knowledge using two complementary approaches. First, we fit an Additive Factors Model (AFM) that predicts step correctness based on prior practice opportunities on each KC \cite{cen2006learning}. Using the fitted model, we estimated each student’s knowledge at their final opportunity per KC and aggregated these estimates into total and average mastery measures. This approach was replicated using the tutor's Bayesian Knowledge Tracing (BKT) estimates. To avoid confounding, this analysis was conducted with the refined KC model applied to both conditions. All process analyses are reported as exploratory, as the preregistered analysis focused on test-score learning gains. Exploratory linear mixed-models adjusted for repeated measures across students, though we also confirmed the robustness of results after additionally adjusting for class periods.

\subsection{Results: Effect on Students’ Learning Gains (RQ2)}
The main effect of test time was significant, indicating substantial pre-to-post learning gains across conditions ($t(279) = 6.14$, $p < .001$). Both conditions exhibited comparable improvements: students using the Original Tutor increased from $M = 0.41$ ($SD = 0.34$) to $M = 0.63$ ($SD = 0.34$), corresponding to a mean gain of $0.21$ ($SD = 0.38$, $d = 0.57$), while students using the Redesigned Tutor improved from $M = 0.39$ ($SD = 0.33$) to $M = 0.59$ ($SD = 0.36$), yielding a mean gain of $0.20$ ($SD = 0.31$, $d = 0.64$). Learning gains did not differ significantly by condition ($t(279) = -0.30$, $p = .760$), nor by unit or their interaction ($ps > .20$).

\subsection{Results: Effect on Students’ Learning Processes (RQ3)}

\paragraph{Manipulation Checks}

Our theory of change predicts that the Redesigned Tutor reallocates practice toward previously hidden, split KCs. It further predicts greater overall difficulty due to prioritizing harder KCs \cite{huang2023supporting}. To test these predictions, we extended the AFM with a condition effect. Practice was significantly more difficult in the Redesigned condition ($\beta = -0.10$, $p = .020$). 
A linear mixed-effects model of practice opportunities showed that students in the Redesigned condition received more opportunities overall ($\beta = 0.30$, $p < .001$). Although new KCs received fewer opportunities than established KCs ($\beta = -0.74$, $p < .001$), the Redesigned condition allocated relatively more practice to them ($\beta = 0.23$, $p = .041$). There was no reliable evidence that the redesign reduced overpractice beyond 80\% mastery, as neither the main effect ($\beta = 0.16$, $p = .080$) nor the interaction with KC type ($\beta = 0.28$, $p = .083$) reached significance.

\paragraph{Time-on-task.}

Students in the Redesigned Tutor, despite equivalent session durations, spent more time logged into the ITS ($M = 47.7$, $SD = 25.7$) than those in the Original Tutor ($M = 39.3$, $SD = 27.0$), $t(235) = 2.44$, $p = .015$. They also engaged in more active, non-idle time ($M = 38.9$, $SD = 21.8$) compared to the Original Tutor ($M = 31.6$, $SD = 20.4$), $t(234) = 4.62$, $p < .001$. Idle time was comparable across conditions (Redesigned: $M = 8.80$, $SD = 17.5$; Original: $M = 7.70$, $SD = 15.8$), with no significant difference, $t(234) = 0.46$, $p = .650$, indicating that increased logged-in time reflects greater productive engagement.

\paragraph{Productivity and unit-level differences.}
Next, we compared productivity between conditions, measured as the number of completed problem steps. In a mixed-effects model predicting the number of completed steps, the main effect of condition was not significant ($\beta = -2.73$, $p = .699$). However, exploratory analyses revealed a significant condition-by-unit interaction such that in Unit~7.06 (Graphs and Equations), where students in the Redesigned Tutor completed more steps than those in the Original Tutor ($\beta = 37.76$, $p = .001$) (but not in other units). In this unit, differences were most pronounced in logged-in time: relative to the Original Tutor, the Redesigned Tutor showed a $28\%$ increase in Unit~7.06, compared to a $16\%$ increase in Unit~7.05, a $4\%$ decrease in Unit~1 Equations, and an $18\%$ increase in Unit~2 Equations.

\paragraph{Knowledge mastered.}
Based on iAFM estimates, students mastered more knowledge in the Redesigned condition (M = 7.61, SD = 2.88) than in the Original condition (M = 6.03, SD = 2.26), $t(115) = 4.33$, $p < .001$, $d_p = 0.40$. A similar pattern was observed using BKT-based estimates (Redesigned: M = 8.96, SD = 3.44; Original: M = 7.30, SD = 2.53), $t(115) = 3.67$, $p < .001$, $d_p = 0.34$. Importantly, this increase reflected broader coverage of skills (since the Redesigned condition included five more skills than the Original condition; see Section \ref{sec:redesign-process-general}). Average KC mastery did not differ significantly between conditions using either AFM (0.683 vs. 0.658, $t(115) = 1.11$, $p = .270$) or BKT estimates (0.687 vs. 0.672, $t(115) = 0.48$, $p = .630$).

\section{Discussion} 

Data from instructional technology often inspires how to improve it. We tested the generality of the data-driven instructional redesign process by Huang et al. \cite{huang2021general}. Our study examined whether the application of the framework could yield interesting discoveries and redesigns, and ultimately improve student learning in four problem sets of an ITS for middle-school mathematics.

The Redesigned Tutor changed students’ learning processes by increasing time-on-task and total knowledge mastered. Although the average mastery level per skill was comparable between conditions, students in the redesigned condition practiced five additional skills within the same class time. This increase can be attributed to greater amounts of completed steps. This was especially evident in Unit 7.06, where reduced complexity likely decreased idle time. The adaptive problem selection algorithm distributed practice across KCs, especially new KCs \cite{huang2021general,xia2025optimizing}. Students also were logged in longer in the redesigned condition, consistent with prior evidence of variability in classroom time use \cite{gurung2025starting}. These differences may reflect motivational factors, a question for future work.

These results are consistent with the idea that making difficult or implicit KCs explicit and supporting them with targeted practice improves learning efficiency \cite{Liu2017ExplanatoryModels,Koedinger2013StudentModels,huang2021general}. Compared to past work on ``focused practice'' approaches, our redesign contributes positive evidence for novel scalable mechanisms such as fast-forwarding over mastered steps \cite{xia2025optimizing} and the use of partially completed problems and part tasks, as in Unit 7.06. However, limited classroom time and variable student engagement likely constrained downstream learning gains.

This study provides further evidence that data-driven redesign, following Huang et al. \cite{huang2021general} and centered on KC model refinement, can improve learning outcomes. We applied this approach flexibly across units, adapting to data-driven issues: algebra units emphasized skill splitting and KC refinement, whereas Unit 7.06 focused on task design and interface scaffolding to improve efficiency. This flexibility aligns with prior work highlighting the importance of domain-specific redesign, including KC modeling \cite{Koedinger2013StudentModels} and curricular structure \cite{Lovett2008OLIStatistics}.

Unlike prior work \cite{Liu2017ExplanatoryModels,Koedinger2013StudentModels,huang2021general}, we observed no significant differences in learning gains between the Original and Redesigned tutor, despite differences in estimated mastery. One explanation is limited post-test coverage of newly introduced skills, including the absence of interpolation items for Unit 7.05, which constrains our ability to detect gains. A second, likely factor is limited practice time. Students completed only six 20-minute sessions across four units, substantially less exposure than in prior studies \cite{huang2021general}. Accordingly, students in both conditions failed to reach mastery on average: AFM estimates remained below 70\%. These results suggest that the benefits of the redesign were limited by insufficient practice. %

\subsection{Limitations, Future Work, and Outlook}

Although the redesign process was highly iterative, students had limited opportunities to reach mastery. Longer practice periods and improved methods for estimating time to mastery are therefore needed, potentially by incorporating indicators of disengagement and struggle into simulations. Although this study extends data-driven redesign to a broader set of content, it remains close to prior mathematics and STEM domains \cite{Liu2017ExplanatoryModels,Koedinger2013StudentModels,huang2021general,Lovett2008OLIStatistics}, prompting domain extensions.%

In closing, data-driven redesign aims to optimize educational technologies, yet its implementation and evaluation remain challenging. We examined a data-driven redesign of a middle-school mathematics ITS grounded in KC model refinement and focused practice. Results show that such refinements can improve productive engagement and expand opportunities for skill practice and mastery, while revealing challenges in ensuring sufficient practice allocation to satisfy mastery criteria. These findings illustrate the promise of data-driven redesign and the need for continued refinement to support instructional technologies.

\begin{credits}
\subsubsection{\ackname}
This research was funded by the Institute of Education Sciences (IES) of the U.S. Department of Education (Award \#R305A220386).
\end{credits}

\bibliographystyle{splncs04}
\bibliography{main} %

\end{document}